%% file: B0toInvisiblegamma.tex
\documentclass[aps,prl,twocolumn,superscriptaddress,showpacs,preprintnumbers,amsmath,amssymb]{revtex4-1}

\usepackage{lineno}
\usepackage{gensymb}
\usepackage{graphicx} 
\usepackage{dcolumn}  
\usepackage{multirow} 
\usepackage{tikz}

\usetikzlibrary{shapes,arrows,positioning,automata,backgrounds,calc,er,patterns}
\usepackage{tikz-feynman}
\tikzfeynmanset{compat=1.0.0}
\graphicspath{{ps}}

\begin{document}
\preprint{\vbox{
		\hbox{Belle Preprint 2020-03, KEK Preprint 2019-59}
}}
\title{ \quad\\[1.0cm] Search for $B^0$ Decays to Invisible Final States ($+\gamma$) at Belle}

\input{pub554_authorlist}

\begin{abstract}
We report searches for $B^0\to\rm{invisible}$ and $B^0\to\rm{invisible}+\gamma$ decays, where the energy of the photon is required to be larger than 0.5 GeV. 
These results are obtained from a $711\,{\rm fb}^{-1}$ data sample 
that contains $772 \times 10^6 B\bar{B}$ pairs and was collected 
near the $\Upsilon\,(4S)$ resonance
with the Belle detector at the KEKB $e^+ e^-$
collider. We observe no significant signal for either decay and set upper limits on their branching fractions at 90\% confidence level of $\mathcal{B}\,(B^0\to\rm{invisible}) < 7.8\times10^{-5}$ and  $\mathcal{B}\,(B^0\to\rm{invisible}+\gamma) < 1.6\times10^{-5}$.
\end{abstract}


\maketitle

\tighten

\begin{figure}[htb]
\centering
\begin{minipage}{.25\textwidth}	
	\begin{tikzpicture}[scale=0.8, transform shape]
	\begin{feynman}[vertical=i1 to a1]
	\vertex (i1){\(b\)};
	\vertex [ right = of i1](a1);
	\vertex [ below = of a1](a2);
	\vertex [ left = of a2](i2){\(\bar{d}\)};
	\vertex [ right =of a1 ] (b1);
	\vertex [ right =of a2 ] (b2);
	\vertex [ right=of b1 ] (f1){\(\bar{\nu}\)};
	\vertex [ right=of b2 ] (f2){\(\nu\)};
	\diagram* {
		(i1) -- [fermion] (a1) -- [scalar, edge label=\(W^-\)] (b1) --  [anti fermion] (f1),
		(i2) -- [anti fermion] (a2) -- [scalar, edge label=\(W^+\)] (b2) --  [fermion] (f2),
		(b1) --[fermion, edge label'=\({e^-}\)] (b2),
		(a1) --[fermion, edge label'=\({u,c,t}\)] (a2),	
	};
	\vertex [above right=of i1] (r){\((\gamma)\)};
	\draw [photon] ($(i1)!0.4!(a1)$) -- (r);
	\end{feynman}
	\end{tikzpicture}
\end{minipage}）
\begin{minipage}{.23\textwidth}
	\begin{tikzpicture}[scale=0.7, transform shape]
	\begin{feynman}[vertical=b to c]
	\vertex (b);
	\vertex [ above left = of b](a1);
	\vertex [ below left = of b](a2);
	\vertex [ left=of a1 ] (i1){\(b\)};
	\vertex [left = of a2 ] (i2){\(\bar{d}\)};
	\vertex [ right=of b ] (c);
	\vertex [ below right=of c ] (f1){\(\bar{\nu}\)} ;
	\vertex [ above right=of c ] (f2){\(\nu\)} ;
	\diagram* {
		(i1) -- [fermion] (a1) -- [scalar, edge label=\(W^-\)] (b) -- [boson, edge label'=\(Z\)] (c) -- [anti fermion] (f1),
		(i2) -- [anti fermion] (a2) --  [scalar, edge label'=\(W^+\)] (b),
		(c) -- [fermion](f2),
		(a1) --[fermion, edge label'=\({u,c,t}\)] (a2),		
	};
    \vertex [above right=of i1] (r){\((\gamma)\)};
    \draw [photon] ($(i1)!0.4!(a1)$) -- (r); 
	\end{feynman}
	\end{tikzpicture}
\end{minipage}
\begin{minipage}{0.23\textwidth}
	\begin{tikzpicture}[scale=0.7, transform shape]
	\begin{feynman}[vertical=b to c]
	\vertex (b);
	\vertex [ above left = of b](a1);
	\vertex [ below left = of b](a2);
	\vertex [ left=of a1 ] (i1){\(b\)};
	\vertex [left = of a2 ] (i2){\(\bar{d}\)};
	\vertex [ right=of b ] (c);
	\vertex [ below right=of c ] (f1){\(\bar{\nu}\)} ;
	\vertex [ above right=of c ] (f2){\(\nu\)} ;
	\diagram* {
		(i1) -- [fermion] (a1) -- [fermion, edge label=\({u,c,t}\)] (b) -- [boson, edge label'=\(Z\)] (c) -- [anti fermion] (f1),
		(i2) -- [anti fermion] (a2) --  [anti fermion, edge label'=\({\bar{u},\bar{c},\bar{t}}\)] (b),
		(c) -- [fermion](f2),
		(a1) --[scalar, edge label'=\(W^-\)] (a2),	
	};
    \vertex [above right=of i1] (r){\((\gamma)\)};
    \draw [photon] ($(i1)!0.4!(a1)$) -- (r); 
	\end{feynman}
	\end{tikzpicture}
\end{minipage}
\caption{Feynman diagrams for $B^0\to \,(\gamma)\nu\bar{\nu}$ in the Standard Model.}
\label{fig1}
\end{figure}
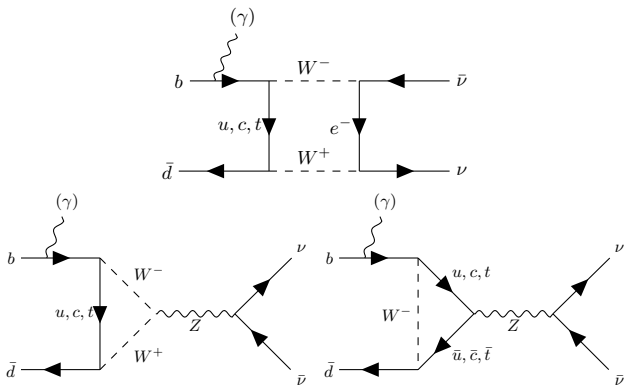

{\renewcommand{\thefootnote}{\fnsymbol{footnote}}}
\setcounter{footnote}{0}
The decays $B^0\to\rm{invisible}$ and $B^0\to\rm{invisible}+\gamma$, with ``invisible'' defined as particles that leave no signal in the Belle detector, are sensitive to new physics beyond the Standard Model (SM). For instance, models with R-parity violation~\cite{01} or dark matter contributions~\cite{02} predict that the branching fraction of $B^0$ decays to an invisible final state could be as high as $10^{-6}$ to $10^{-7}$.  In the SM, such a decay is $B^0\to\,(\gamma)\nu\bar{\nu}$, which proceeds through the Feynman diagrams in Fig.~\ref{fig1}. The $B^0\to\nu\bar{\nu}$ decay is strongly helicity suppressed by a factor of $(m_\nu/ {m_{B^0}})^2$~\cite{03}, and the estimated branching fraction is at the $10^{-25}$ level. A recent calculation~\cite{04} predicts that a $B^0\to\nu\bar{\nu}\nu\bar{\nu}$ decay, which has the same signature as $B^0\to\nu\bar{\nu}$ in the detector, also contributes to the invisible final state, and the estimated branching fraction is at the $10^{-16}$ level. For the $B^0\to\gamma\nu\bar{\nu}$ decay, despite the removal of helicity suppression, the branching fraction predicted from the SM is of order $10^{-9}$~\cite{05}, which is still too small to be observed by current experiments. A very low background from the SM indicates that a signal of $B^0\to\rm{invisible}\,(+\gamma)$ in the current B-factory data would indicate new physics.\par

Several experimental searches for $B^0\to\rm{invisible}\,(+\gamma)$ have been performed and no signal has been observed. The most stringent 
branching-fraction upper limits~\cite{06}, $\mathcal{B}\,(B^0\to\rm{invisible}) < 2.4\times10^{-5}$ and $\mathcal{B}\,(B^0\to\rm{invisible}+\gamma) < 1.7\times10^{-5}$, were provided by the BaBar Collaboration using the semileptonic tagging method and with $424\,{\rm fb}^{-1}$ of data. A previous search~\cite{07} 
from Belle with $606\,{\rm fb}^{-1}$ of data adopted a hadronic tagging method and reported the upper limit, $\mathcal{B}\,(B^0\to\rm{invisible}) < 1.2\times10^{-4}$, a factor of five higher than the BaBar results. Here we report the updated results with the full Belle data set and improved hadronic tagging.
 
These searches are based on a data sample that was collected with the Belle detector at the KEKB asymmetric-energy
$e^+e^-$ (3.5 on 8 GeV) collider~\cite{08}. The sample
contains $772 \times 10^6\, B\overline{B}$ pairs accumulated at the $\Upsilon\,(4S)$ resonance, corresponding to an integrated luminosity of  $711\,{\rm fb}^{-1}$, and an additional $90\,{\rm fb}^{-1}$ of off-resonance data recorded at the center-of-mass (CM) energy 60 MeV below the $\Upsilon(4S)$ resonance.\par

The Belle detector is a large-solid-angle magnetic
spectrometer that consists of a silicon vertex detector (SVD),
a 50-layer central drift chamber (CDC), an array of
aerogel threshold Cherenkov counters,  
a barrel-like arrangement of time-of-flight
scintillation counters, and an electromagnetic calorimeter (ECL)
comprised of CsI (Tl) crystals located inside 
a superconducting solenoid coil that provides a 1.5~T
magnetic field. Outside the coil, the $K_L^0$ and muon detector (KLM), composed of alternating layers of charged particle detectors and iron plates, is instrumented to detect $K_L^0$ mesons and to identify
muons. The detector is described in detail elsewhere~\cite{09}.
Two inner detector configurations were used. A 2.0 cm radius beampipe
and a 3-layer SVD were used for the first $140\,{\rm fb}^{-1}$ data sample, while a 1.5 cm radius beampipe, a 4-layer 
SVD and a small-cell inner CDC were used to record  
the remaining $571\,{\rm fb}^{-1}$ data sample~\cite{10}. \par

To determine our signal efficiency and optimize event selection criteria, we use Monte Carlo (MC) simulated events. All MC samples in the analysis are generated by the \textsc{EvtGen} package~\cite{11}, with the detector response simulated by the \textsc{Geant3} package~\cite{12}. Ten million $B^0\to\nu\bar{\nu}$ and $B^0\to\gamma\nu\bar{\nu}$ signal events are generated with a phase-space decay model. However, for the $B^0\to\gamma\nu\bar{\nu}$ search, a phase-space decay model is not appropriate to describe the process. Thus, the signal efficiency is reweighted according to theoretical calculations~\cite{05}, in which the ``quark constituent model" is assumed and differential branching fraction as a function of squared missing mass (${M^2_{\rm{miss}}}$) is given. ${M^2_{\rm{miss}}}$ is defined as:
 \begin{equation}
{M^2_{\rm{miss}}} = \,(\vec{P}_{\rm{beam}}-\vec{P}_{B_{\rm{tag}}}-\vec{P}_\gamma)^2/c^2,
\label{miss}
 \end{equation}
 where $\vec{P}_{\rm{beam}}$, $\vec{P}_{B_{\rm{tag}}}$ and $\vec{P}_\gamma$ are the four-momenta of the $e^+e^-$ system, the other $B$ meson and the photon for a $B^0\to\,\gamma\nu\bar{\nu}$ signal event, respectively. In addition, a second model-independent binned analysis is performed in five different ${M^2_{\rm{miss}}}$ regions using the signal MC sample generated with the phase-space decay model: ${M^2_\text{miss}}<5$ GeV$^2/c^4$, $5\text{ GeV}^2/c^4<{M^2_\text{miss}}<10$ GeV$^2/c^4$, $10\text{ GeV}^2/c^4<{M^2_\text{miss}}<15$ GeV$^2/c^4$, $15\text{ GeV}^2/c^4<{M^2_\text{miss}}<20$ GeV$^2/c^4$ and $20\text{ GeV}^2/c^4<{M^2_\text{miss}}$ (bin1-bin5, respectively). \par
 
Since the signal-side particles, except for the photon, cannot be detected, a technique that fully reconstructs the other $B$ meson (tag-side $B_{\rm{tag}}$ meson) is used. The signature of $B^0\to\rm{invisible}$ or a photon for $B^0\to\rm{invisible}+\gamma$ is then identified in the remaining part of the event.\par

The hadronic full reconstruction is a hierarchical process for reconstructing the $B_{\rm{tag}}$ meson~\cite{13}.
The $B^0$ candidates are reconstructed from 489 decay channels in which $B^0$ mesons decay to hadrons. The process consists of four stages, starting from an initial selection of charged tracks, photons, $K_S^0$, and $\pi^0$, followed by two stages of forming intermediate particles, ($D^\pm_{(s)}$, $D^0$, $J/\psi$) and ($D^{*\pm}_{(s)}$, $D^{*0}$), and ending at the stage of reconstructing the $B^0$ meson from its daughter products. The neural network (NN) package, \textsc {NeuroBayes}~\cite{14}, is used to assign a signal probability ($P_{\rm FR}$) to the reconstructed particle at each individual stage. The NN at each stage is trained with the $P_{\rm FR}$ of the daughter particles and properties of the candidate, such as invariant mass and the opening angle between daughters. If there are multiple $B^0$ meson candidates in an event, the candidate with the highest $P_{\rm FR}$ is selected as the $B_{\rm{tag}}$. From the previous study~\cite{13}, the number of correctly reconstructed $B_{\rm{tag}}$ in the full data set is $1.4\times10^6$. In the case of $B^0\to\nu\bar{\nu}$ and $B^0\to\gamma\nu\bar{\nu}$ signal MC simulation, the reconstruction efficiencies of the $B_{\rm{tag}}$ are $0.41\%$ and $0.47\%$, respectively. Comparing to the full reconstruction algorithm used in the previous $B^0\to\rm{invisible}$ study at Belle~\cite{07}, the tagging efficiency is improved by approximately a factor of $1.5$ due to the newly introduced NN tool within the framework. In this analysis, a loose preselection on the beam-energy-constrained $B_{\rm{tag}}$ mass, $M_{\rm bc,tag}>5.26\text{ GeV}/c^2$, is applied. This mass is calculated as $M_{\rm bc,tag}=\sqrt{E^2_{\rm{beam}}-\vec{P}^{\, 2}_{B_{\rm{tag}}}c^2}/c^2$, where the $E_{\rm beam}$ is the beam energy in the $e^+e^-$ CM frame, and the $\vec{P}_{B_{\rm{tag}}}$ is also defined in this frame.\par  

For $B^0\to\rm{invisible}+\gamma$, at least one photon is required. The signal photon is detected by the ECL and an energy threshold of 0.5 GeV in the $e^+e^-$ CM frame is applied in order to eliminate the huge number of photons from the beam background. Furthermore, we require that the corresponding ECL cluster does not match with a track in the CDC, and that the fraction of energy detected in the inner $3\times3$ array of crystals relative to the $5\times5$ array of crystals centered on the crystal with the maximum energy exceeds 0.9. In the case that more than one photon satisfies the selection criteria, the one with the highest energy is selected as the signal photon.\par

After the reconstruction of $B_{\rm{tag}}$, and selecting the photon for $B^0\to\rm{invisible}+\gamma$, events with extra tracks, $\pi^0$, or $K_L^0$ are rejected because no extra detectable particles except photons are expected in the event. Extra tracks are defined as those passing the loose impact parameter selections $dr<4\text{ cm}$ and $|dz|<35\text{ cm}$, where $dr$ and $dz$ are the shortest distance from the track to the interaction point (IP) on the transverse plane and along the beam axis, respectively. The loose requirement aims to include low-momentum tracks that are ill-reconstructed and tracks not produced around the IP. Extra $\pi^0$ candidates are reconstructed from photon pairs passing the following requirements: each photon has energy larger than 40 MeV; the absolute cosine value of the angle between a photon direction and the boost direction of the lab system in the $\pi^0$ rest frame smaller than 0.9; $120\text{  MeV}/c^2< M_{\pi^0}<145$ MeV$/c^2$, which corresponds to a window within 1.5 standard deviations ($\sigma$) of the nominal mass~\cite{15}. Extra $K_L^0$ candidates are detected in the KLM detector, where a minimum of two hit layers is required. \par 

A powerful variable to identify $B^0\to\rm{invisible}$ and $B^0\to\rm{invisible}+\gamma$ signal is  $E_{\rm{ECL}}$, which is defined as the sum of all the remaining energies of ECL clusters that are not associated with tag-side $B$ daughter particles. For $B^0\to\rm{invisible}+\gamma$, the signal photon is also excluded. In the $E_{\rm{ECL}}$ calculation, in order to reduce a contribution from beam background, only the ECL clusters that satisfy the following energy thresholds are included: $E_\text{cluster}>0.05$,  $0.10$ and $0.15$ GeV for the barrel region ($32.2\degree<\theta<128.7\degree$), forward endcap ($\theta<32.2\degree$) and backward endcap ($\theta>128.7\degree$), respectively, where $\theta$ is the polar angle in the lab frame. Since the distribution for signal events peaks at zero, the $E_{\text{ECL}}$ signal box is defined as $E_{\text{ECL}}<0.3\,\text{GeV}$, and the $E_{\text{ECL}}$ sideband is defined as $0.3\,\text{GeV}<\,E_{\text{ECL}}<1.2\,\text{GeV}$.  \par

\begin{figure}[htb]
 	\includegraphics[width=0.48\textwidth]{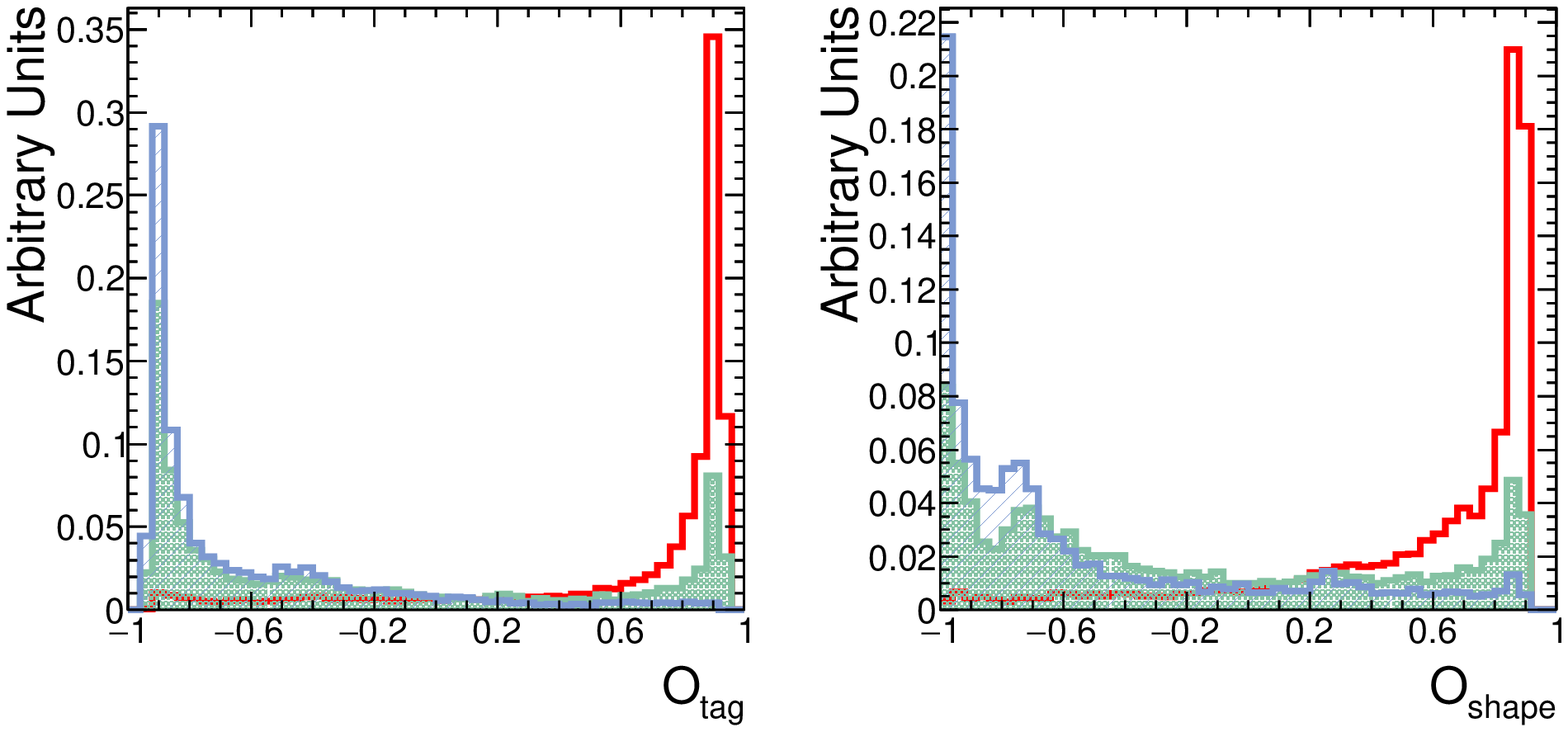}
 	\includegraphics[width=0.48\textwidth]{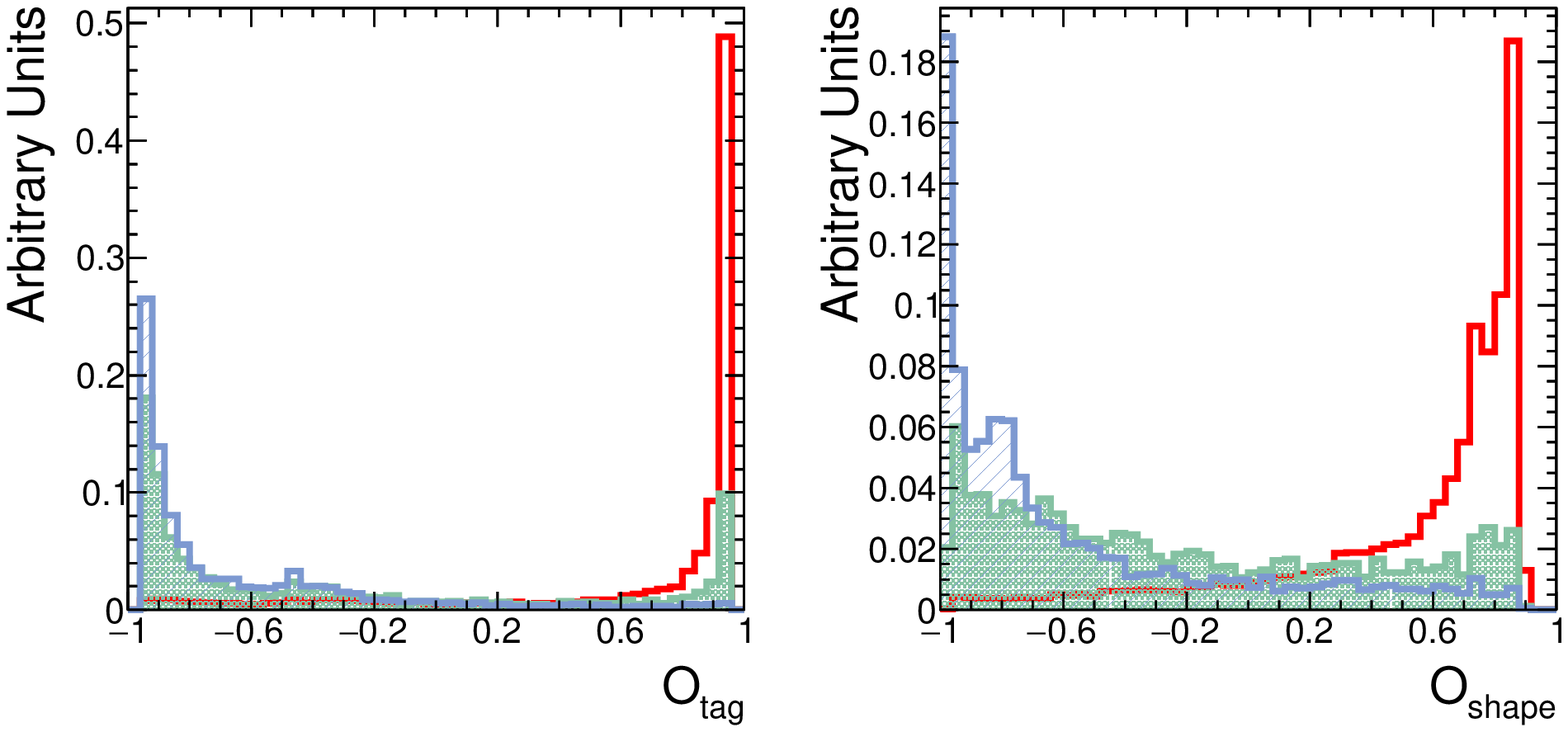}
 	\caption{$O_\text{tag}$ (left) and $O_\text{shape}$ (right) distributions for continuum (blue and hatched area), generic $B$ (green and shaded area) background, and signal MC simulation (red and blank area). Top: $B^0\to\text{invisible}$; Bottom: $B^0\to\text{invisible}+\gamma$. Histograms are normalized such that the sum of all bins equals one.}
 	\label{fig2}
 \end{figure}

After the signal event selections, $e^+e^-\to q \bar{q}\,(q=u,d,s,c)$ continuum events are the dominant background, followed by $B\bar{B}$ decay with a $b\to c$ transition (generic $B$ background). Two separate NN implemented using the \textsc{NeuroBayes} package are used in order to reduce the former. The first NN focuses on rejecting fake $B_{\text{tag}}$, the input variables are those related to the $B_{\text{tag}}$ reconstruction qualities: $P_\text{FR}$ of the $B_\text{tag}$; $M_{\rm bc,tag}$; $\Delta E_{\rm tag}$, which is defined as the energy difference between the reconstructed $B_{\rm tag}$ meson and the beam energy at the $e^+e^-$ CM frame. The second NN focuses on the jet-like topology of continuum events. The input variables are the sum of the transverse momentum, $M_\text{miss}^2$, which is defined in Eq.~\ref{miss} without the $\vec{P}_\gamma$ term, and sixteen modified Fox-Wolfram moments~\cite{16}. For $B^0\to\text{invisible}+\gamma$, the signal photon is excluded in all the momentum-related calculations in order to reduce model dependence. Outputs of the two NN ($O_\text{tag}$ and $O_\text{shape}$ respectively) are continuous variables within the range $(-1,1)$, and larger (smaller) values correspond to events more (less) likely to be signal. We find that $O_\text{tag}$ and $O_\text{shape}$ are also effective at distinguishing the generic $B$ background from the signal. The $O_\text{tag}$ and $O_\text{shape}$ distributions for signal and both kinds of the background are shown in Fig.~\ref{fig2}. \par

  Thresholds for $O_\text{tag}$ and $O_\text{shape}$ are determined jointly by maximizing a figure of merit (FOM) separately for the modes, $B^0\to\text{invisible}$ and $B^0\to\text{invisible}+\gamma$, and the five $M_\text{miss}^2$ bins. The optimization is done in the $E_\text{ECL}$ signal box and the FOM has the form~\cite{17}:
  \begin{equation}	\text{FOM}=\frac{\epsilon_\text{sig}}{(0.5n_\sigma+\sqrt{N_\text{bkg}})}, 
  \end{equation}
   where $\epsilon_\text{sig}$ is the signal efficiency in MC simulation and $N_{\text{bkg}}$ is the number of background events reconstructed as signal in MC. Here $n_\sigma$ is the number of $\sigma$ in a one-tailed Gaussian test, where $n_\sigma=1.28$ corresponds to the choice of a 90\% confidence level. The optimized NN output thresholds, $O_\text{tag}>0.7$ and $O_\text{shape}>-0.1\,(-0.2)$ for $B^0\to\text{invisible}$ ($B^0\to\text{invisible}+\gamma$), eliminate 97\% of background events while retaining around 60\% of signal in both cases. For different bins in the binned analysis, lower bounds for the $O_\text{tag}$ and $O_\text{shape}$ lie between $(0,0.7)$ and $(-0.4,0.2)$, respectively. With the thresholds, 92\%---98\% of background events are reduced while 60\%---80\% of signal events are kept.   \par
   
   The signal yield for $B^0\to\text{invisible}$ is extracted from data through fitting variables $E_{\text{ECL}}$ and $\cos\theta_T$, where $\cos\theta_T$ is the cosine of the angle between the two thrust axes in the $e^+e^-$ CM frame. The thrust axis is defined as the direction that maximizes the sum of the longitudinal momenta of particles, and here one of the axes is constructed using $B_{\text{tag}}$ final-state particles, while the other is from the remaining part of the event. The latter is composed of photons and charged tracks that survive the extra-track rejection. In case there is no particle in the remaining part, the beam axis replaces the second thrust axis. In data and the signal MC simulation, this occurs in less than $1\%$ of events.\par
   
   Beside generic $B$ and continuum backgrounds, background from rare $B\bar{B}$ decays (i.e., with a $b\to u$, $b\to d$, or $b\to s$ transition) and from $e^+e^-\to\tau^+\tau^-$ are also considered. From MC simulation, it is found that the rare $B\bar{B}$ decay background has $\cos\theta_T$ and $E_{\text{ECL}}$ distributions similar to those of generic $B$ background, and thus those two background sources are combined. In addition, the continuum and $e^+e^-\to\tau^+\tau^-$ background also have similar $\cos\theta_T$ distributions, and their $E_{\text{ECL}}$ combined distribution can be described by the off-resonance data. As a result, continuum and $e^+e^-\to\tau^+\tau^-$ backgrounds are combined and referred to as non-$B$ background.  \par
   
    An extended unbinned maximum likelihood fit is applied with the form:
  \begin{equation}   
   \mathcal{L}=\frac{e^{-\sum_j{n_j}}}{N!}\prod_{i=1}^{N}\,(\sum_{j}n_jP_j\,(E_\text{ECL}^i,\cos\theta_T^i)),
  \end{equation}
  where $i$ is the events identifier, $n_{j}$ is the number of event belonging to the $j$th category. $P_j\,(E_\text{ECL},\cos\theta_T)$ is a direct product of the probability density functions (PDFs) $P_j\,(E_\text{ECL})$ and $P_j\,(\cos\theta_T)$. With the exception that the $E_\text{ECL}$ distribution for the non-$B$  component is obtained from the off-resonance data, all the other PDFs are obtained from MC simulation. In order to enhance the statistics when constructing PDFs, the $O_\text{tag}$ threshold is removed after verifying that no correlation exists between $O_\text{tag}$ and the fitting variables. From the MC simulation, the proportions of the continuum background among the non-$B$ background are $(83\pm5)\%$ and $(75\pm1)\%$ before and after removing the $O_\text{tag}$ threshold, respectively, consistent within $1.6\sigma$ uncertainty. Second-order Legendre polynomials are used to describe $\cos\theta_T$, while histogram PDFs are used for the $E_{\text{ECL}}$ distributions. No correlation is found between the fitting variables in background components. However, a small but nonnegligible correlation between variables exists for signal events. The direct product between PDFs is used nonetheless, and the corresponding systematic uncertainty is determined by generating an ensemble according to two-dimensional histogram PDFs, and then fitting with the product of one-dimensional PDFs.\par 
  
\begin{figure}[htb]
    \includegraphics[width=0.235\textwidth]{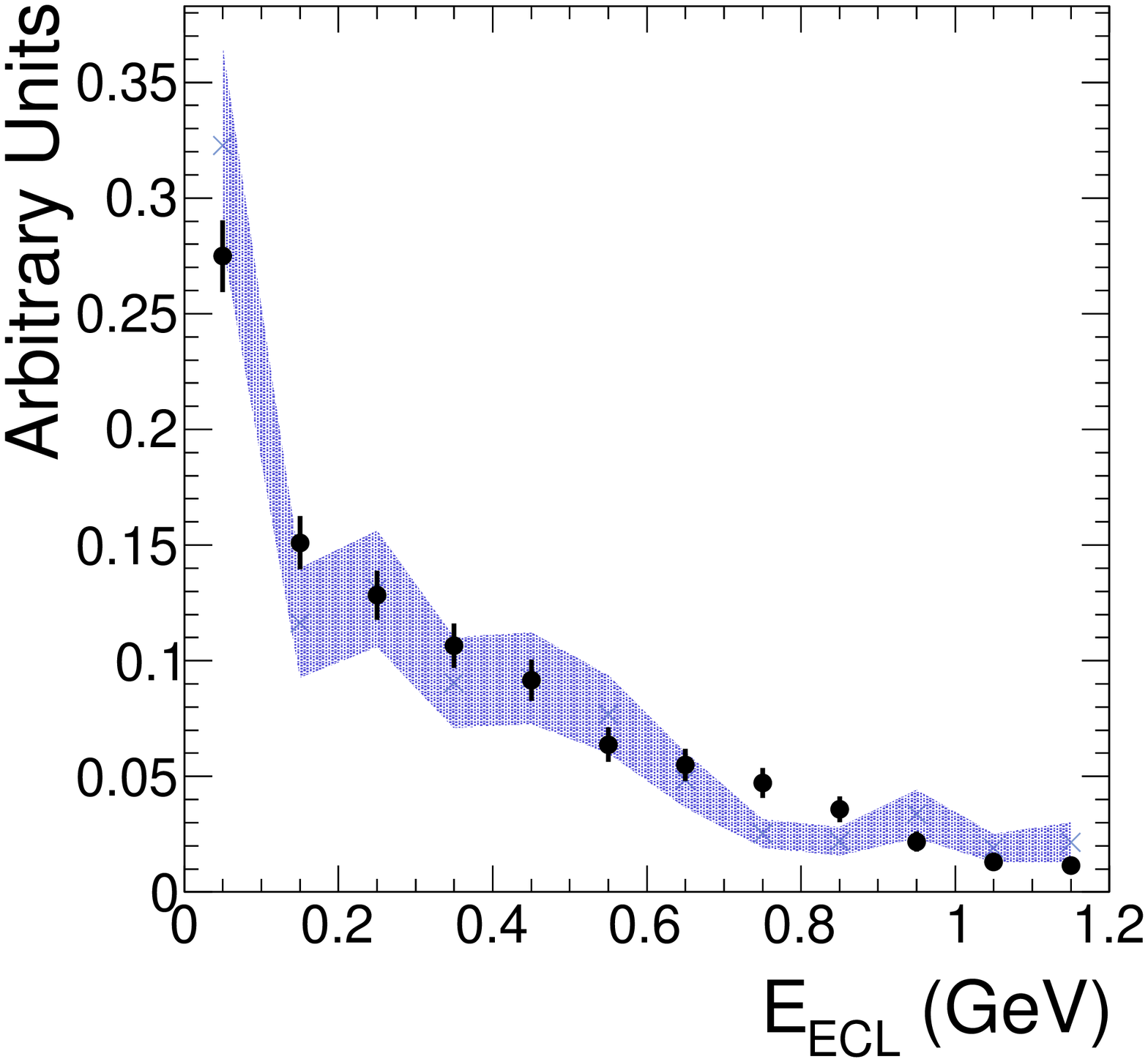}	
  	\includegraphics[width=0.235\textwidth]{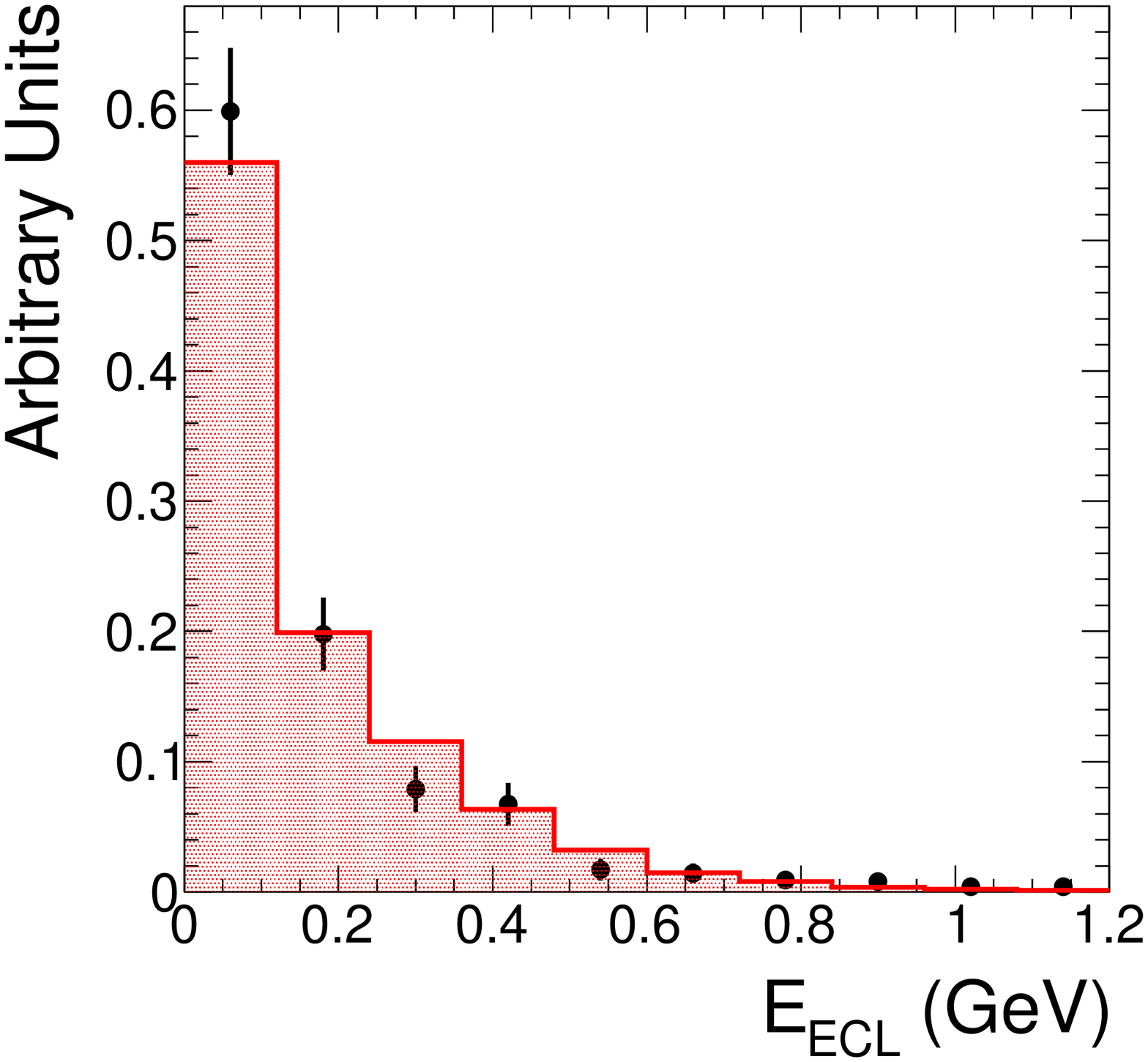}
  	\caption{Left: Comparison of $E_{\text{ECL}}$ distributions between background MC simulation and data in the $O_\text{tag}$ sideband. The black points are data. The blue crosses with a shaded error band are the background MC simulation. Right: Comparison of $E_{\text{ECL}}$ distributions between $B^0\to \nu\bar{\nu}$ signal MC simulation and $B^0\to D^{*-}l^+\nu$ data. The black points are data. The red and shaded distribution is signal MC simulation. Histograms are normalized such that the sum of all bins equals one.}
  	\label{fig3}
    \end{figure}

\begin{figure}[htb]
	\includegraphics[width=0.5\textwidth]{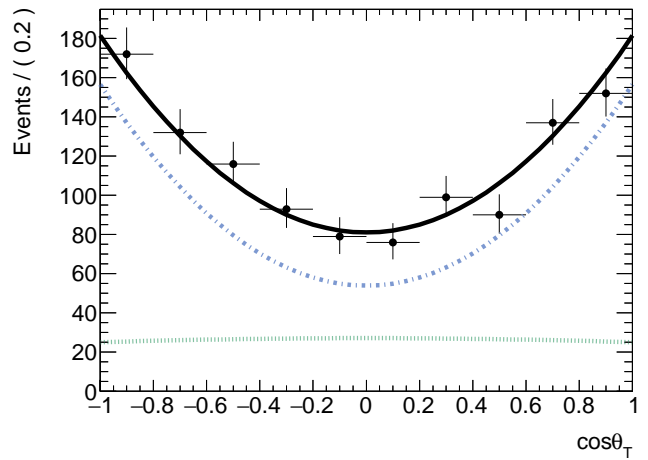} 	
	\caption{Verification of the background ratio in MC simulation in the $O_\text{tag}$ sideband. Dots with error bars are data, black-solid line is the fit result, green short-dashed line is the generic $B$ background component and blue dashed-dotted line is the non-$B$ background component.}
	\label{fig4}
    \end{figure}

  The validity of the $E_{\text{ECL}}$ PDFs for background is checked using the sideband samples excluded by the $O_\text{tag}$ threshold. Comparison between sideband data and the combined distribution of non-$B$ and generic $B$ background according to the MC ratio shows consistency, as shown in Fig.~\ref{fig3}. In the comparison, the correctness of the MC ratio between background components is further verified by fitting $\cos\theta_T$ in the sideband sample, which is shown in Fig.~\ref{fig4}. In this fit, there are $(23\pm8)\%$ of generic $B$ events among the combined background, which is consistent with the proportion of $(25\pm1)\%$ from MC simulation. \par
  
  To verify the $E_{\text{ECL}}$ PDF obtained from the signal MC simulation, $B^0\rightarrow D^{*-}l^+\nu$ ($l=e,\mu, D^{*-}\to \bar{D^0}\pi^-$, $\bar{D^0}\rightarrow K^+ \pi^-$) is used as a control sample. In these events, $B_\text{tag}$ is fully reconstructed, and the other $B$ meson is identified by decays to $D^{(*)}l\nu$ from the remaining part of the event (double tagging). To mimic the invisible final state, particles used in the signal-side reconstruction are excluded, such as in the $E_{\text{ECL}}$ and the shape variables calculations. Event selections are done in the same manner as in the $B^0\to\text{invisible}$ study. The extra tracks, $\pi^0$, and $K_L^0$ vetoes are demanded after removing particles involved in the reconstruction of $B_{\text{tag}}$ and $B_\text{sig}$. The $O_\text{tag}$ and $O_\text{shape}$ are also based on the algorithms established before. Additional selections include: 1.855 GeV$/c^2<M_{D^0}<$1.885 GeV$/c^2$ ($1.8 \sigma$ window); 0.143 GeV$/c^2<\Delta M_D <$0.148 GeV$/c^2$ ($2.2 \sigma$ window), where $\Delta M_D$ is the difference between the reconstructed $D^{*-}$ and $\bar{D^{0}}$ masses; $-0.5\text{ GeV}^2/c^4<M_\text{miss}^2<0.5\text{ GeV}^2/c^4$ ($1.5 \sigma$ window), where $M_\text{miss}^2$ is defined in Eq.~\ref{miss} with $\vec{P}_\gamma$ replaced by $\vec{P}_{D^{*-}l}$. After the double tagging, background for the $B^0\rightarrow D^{*-}l^+\nu$ becomes negligible. Comparison of the $E_{\text{ECL}}$ distribution between the doubly tagged data and the $B^0\to \nu \bar{\nu}$ MC simulation shows excellent agreement as seen in Fig.~\ref{fig3}. \par
  
\begin{figure}[htb]
	\includegraphics[width=0.235\textwidth]{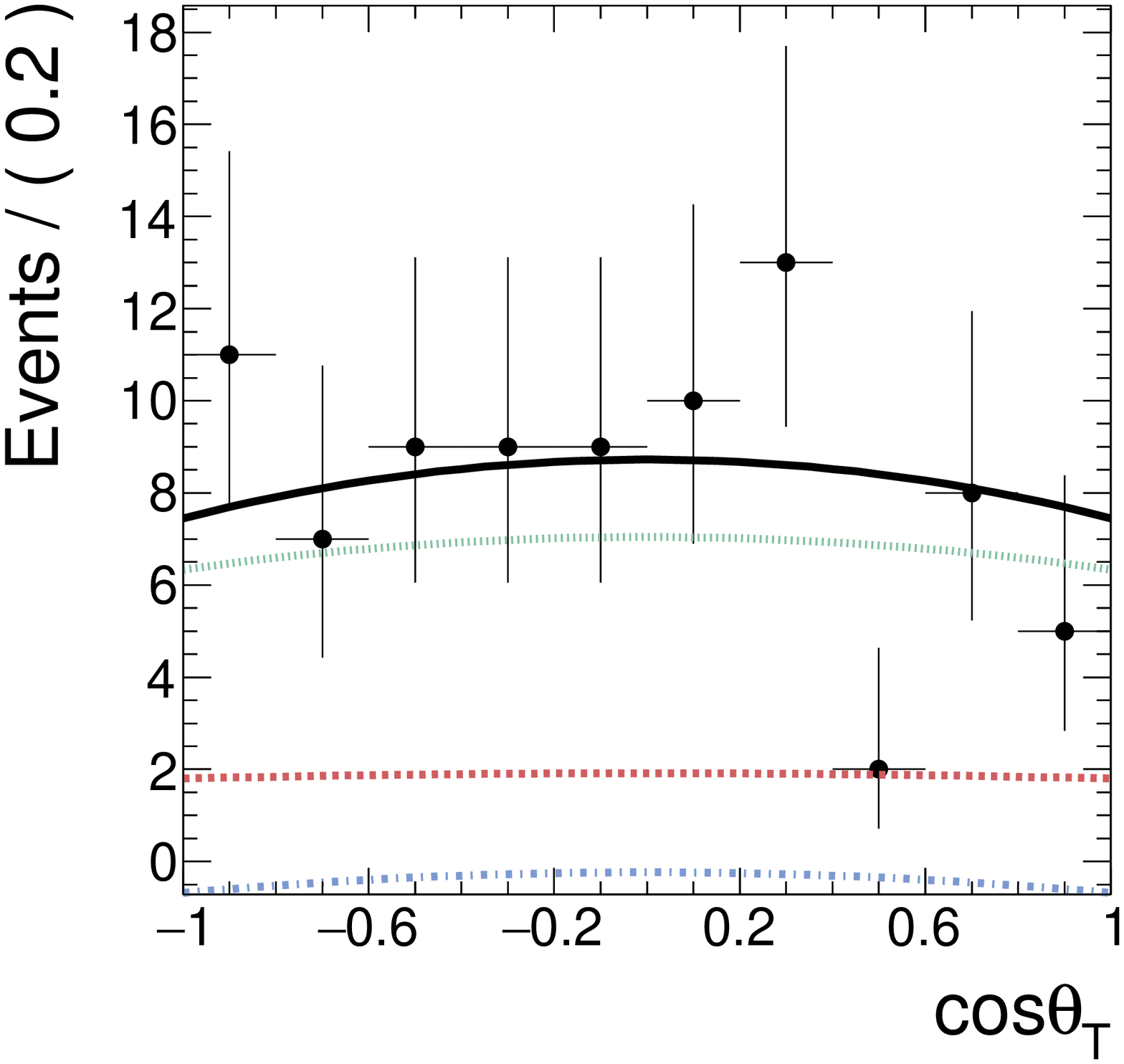}
	\includegraphics[width=0.235\textwidth]{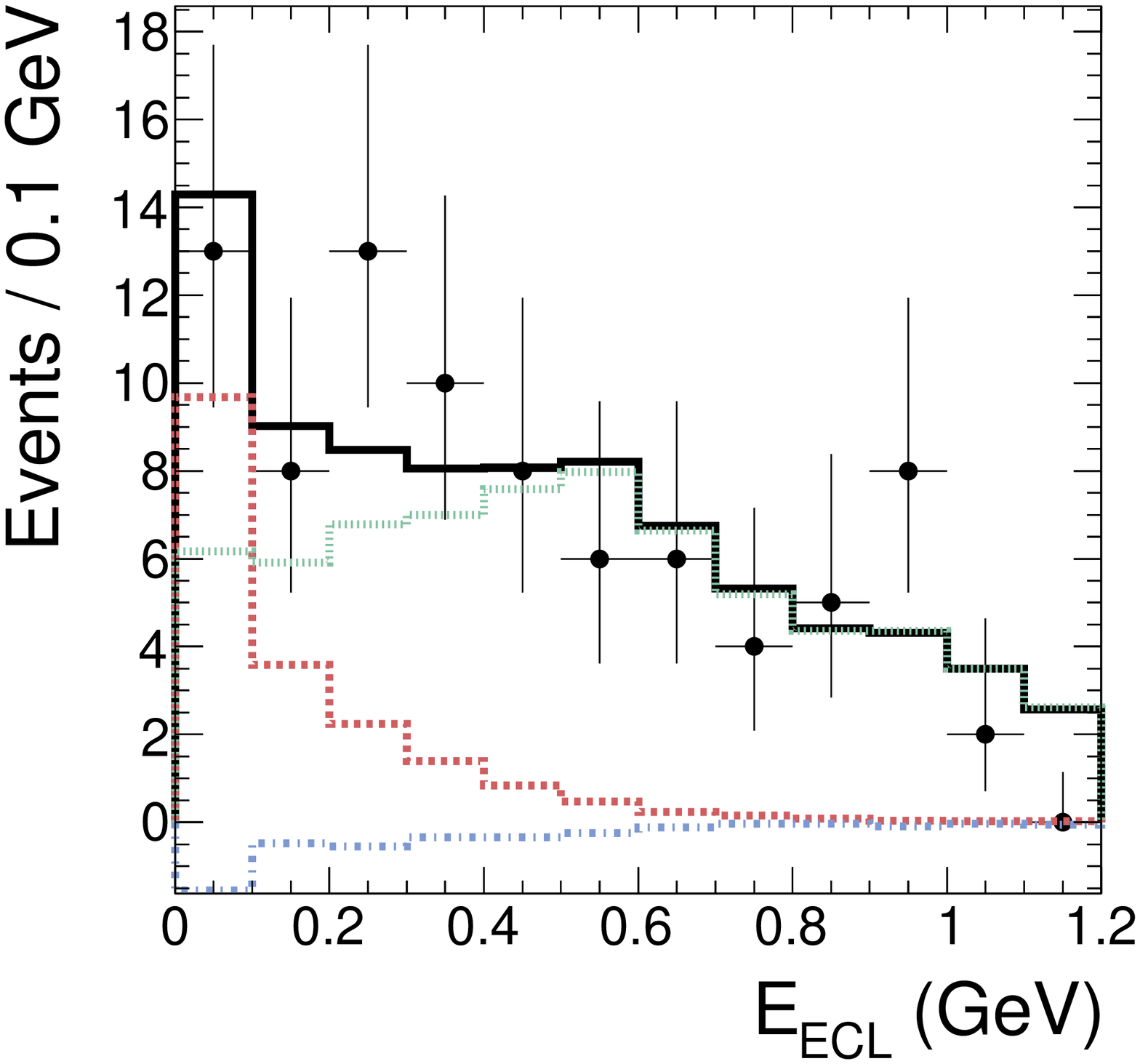}
	\caption{Projections of the fit result on $\cos\theta_T$ (left) and $E_{\text{ECL}}$ (right) for $B^0\to\text{invisible}$. Dots with error bars are data, black-solid line is the fit result, red-dotted line is the signal component, green short-dashed line is the generic $B$ background component and blue dashed-dotted line is the non-$B$ background component.
	}
	\label{fig5}
    \end{figure}

    The projections of the 2D fitting result for $B^0\to\text{invisible}$ are shown in Fig.~\ref{fig5}. The corresponding fitting yields of each component are listed in Table~\ref{Table1}. No significant signal is observed. \par 
    
    \begin{table}[htb]
	\caption{Fitting yield ($B^0\to \text{invisible}$).} 		
	\begin{tabular}
		{@{\hspace{0.5cm}}l@{\hspace{0.5cm}}  @{\hspace{0.5cm}}c@{\hspace{0.5cm}}}	
		\hline\hline
		Component&Yields \\
		\hline
		Signal& $18.8\substack{+15.3 \\ -14.5}$ \\
		Generic $B$&$68.1\substack{+12.2 \\ -11.7}$ \\
		Non-$B$&$-3.9\substack{+19.5 \\ -17.5}$ \\
		\hline\hline		
	\end{tabular}	
	\label{Table1}	
    \end{table}

    The systematic uncertainty due to the statistical error of the $E_{\text{ECL}}$ and $\cos\theta_T$ PDFs modelling is estimated by varying the content of each bin in the histogram PDFs and parameters of the Legendre polynomials by $\pm1\sigma$ respectively and repeating the fit on data. All of the systematic uncertainties of signal yields are listed in Table~\ref{Table2}, and the total systematic uncertainty is the sum in quadrature of all terms.\par
    
    \begin{table}[htb]
  	\caption{Summary of systematic uncertainties on fitting yield.}
  	\label{Table2}
	\begin{tabular} {@{\hspace{0.5cm}}l@{\hspace{0.5cm}}  @{\hspace{0.5cm}}c@{\hspace{0.5cm}}}
	\hline\hline
	Sources & Sys. uncertainty (Events)\\
	\hline
	Signal PDF & $\pm 0.6$\\
	\multirow{2}{*}{Generic $B$ PDF} & $+1.9$\\
	& $-1.8$\\                
	\multirow{2}{*}{Non-$B$ PDF} & $+6.6$\\
	& $-6.7$\\     	
	\multirow{2}{*}{Signal PDF correlation} & $+0.3$\\
	& $-0.0$\\
	\hline
	\multirow{2}{*}{Total} & $+6.8$\\
	& $-7.0$\\   	
	\hline\hline 
  \end{tabular}
  \end{table}

 The significance of the signal yield is defined as $\sqrt{-2 \text{ln}(\mathcal{L}_0/\mathcal{L}_s})$, where $\mathcal{L}_0$ and $\mathcal{L}_s$ are the maximized likelihood values when the signal yield is constrained to zero and floated, respectively. The systematic uncertainty is taken into consideration by convolving the likelihood function with a Gaussian function whose width equals to the systematic uncertainty. The signal significance thus obtained for $B^0\to\text{invisible}$ is $1.2\sigma$.\par 
 
 Since few events are expected in data for $B^0\to\text{invisible}+\gamma$ and in the binned analysis, an approach that counts events in the $E_\text{ECL}$ signal region and then subtracts the background is employed to measure any signal. The number of background events in the signal box ($N_{\text{bkg, box}}^{\text{\,data}}$) is estimated from the $E_{\text{ECL}}$ sideband data ($N_{\text{bkg, s.b.}}^{\text{\,data}}$) by multiplying by a parameter $N_\text{bkg, box}^{\text{\,MC}}/N_\text{bkg, s.b.}^{\text{\,MC}}$:
  \begin{equation}
 N_\text{bkg, box}^{\text{\,data}} =\\ 
 N_\text{bkg, s.b.}^{\text{\,data}} \times \frac{N_\text{bkg, box}^{\text{\,MC}}}{N_\text{bkg, s.b.}^{\text{\,MC}}},
 \label{ratio}
  \end{equation}
 where the $N_\text{bkg, box}^{\text{\,MC}}$ and $N_\text{bkg, s.b.}^{\text{\,MC}}$ denote the number of background events in the $E_{\text{ECL}}$ signal box and sidebands from MC simulation. \par

\begin{figure}[htb]
  	\includegraphics[width=0.235\textwidth]{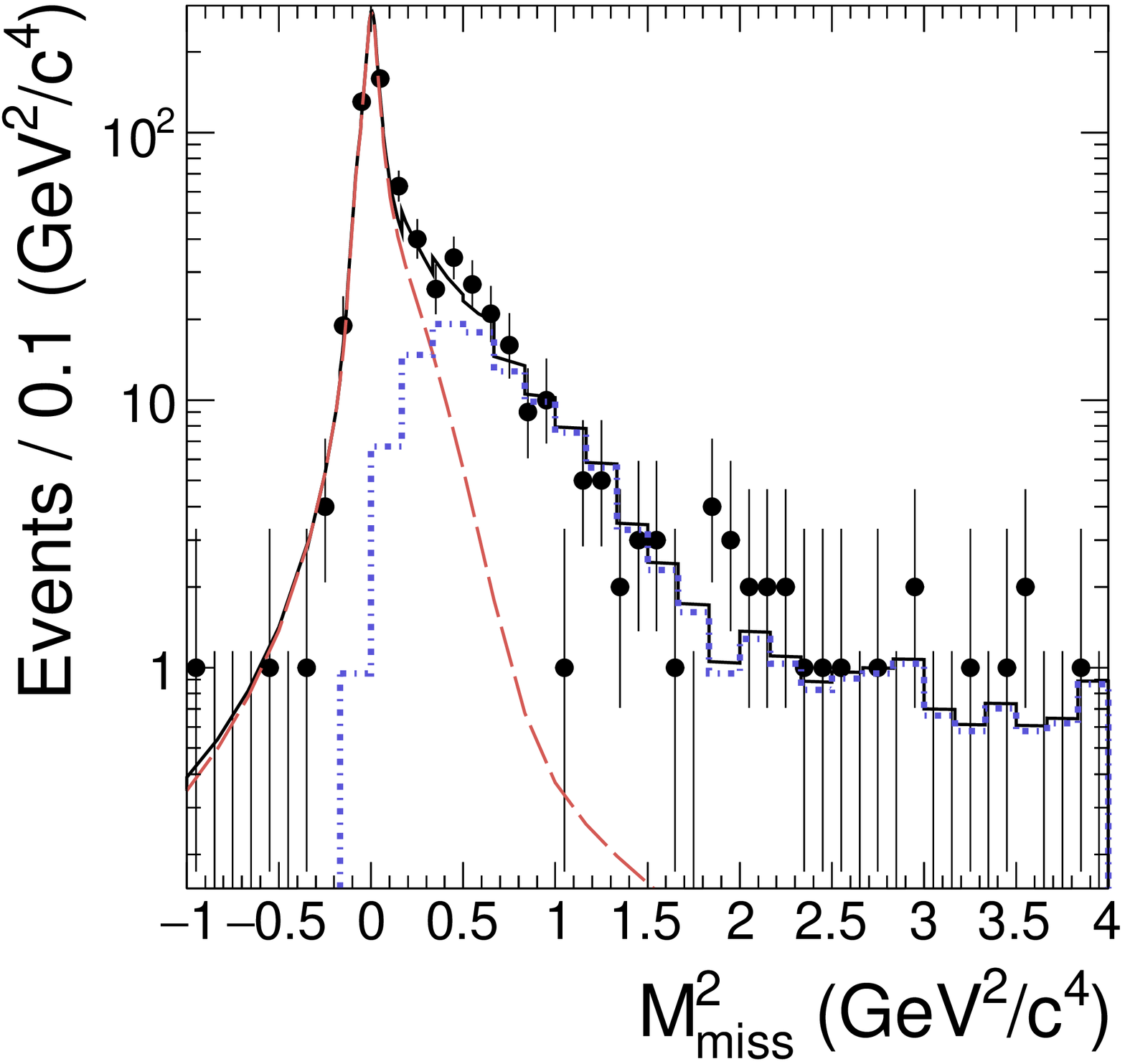} 
  	\includegraphics[width=0.235\textwidth]{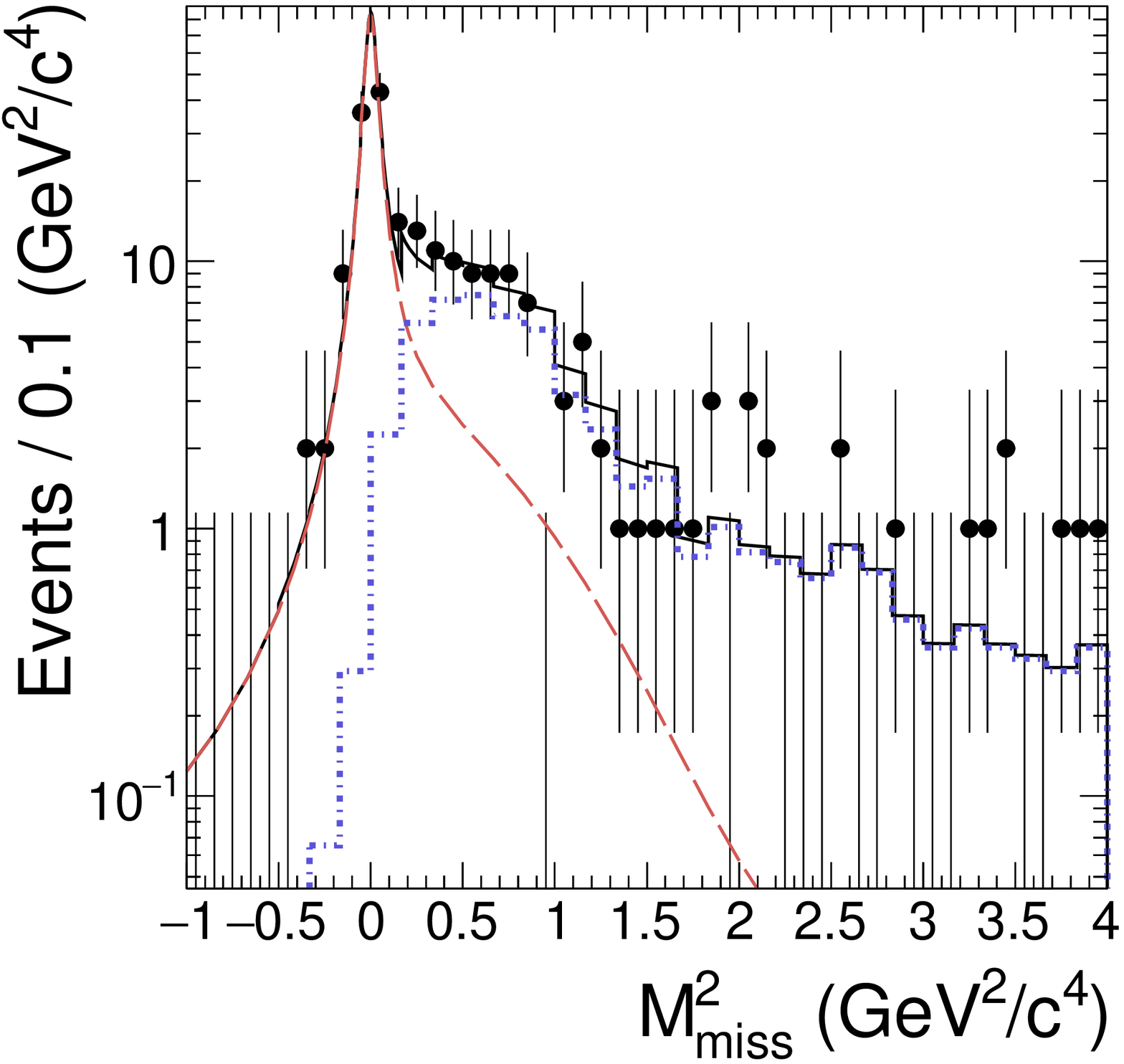}	
  	\caption{Fitting result of the control sample $B^0\to D^-l^+\nu$ in the $E_\text{ECL}$ signal box (left) and sideband (right). Selections are based on $B^0\to\text{invisible}\,+\gamma$. Dots with error bars are data, black-solid line is the combined fit result, red-dashed line is the signal component and blue dashed-dotted line is background.}
  	\label{fig6}
    \end{figure}

\begin{figure}[htb]
	\includegraphics[width=0.235\textwidth]{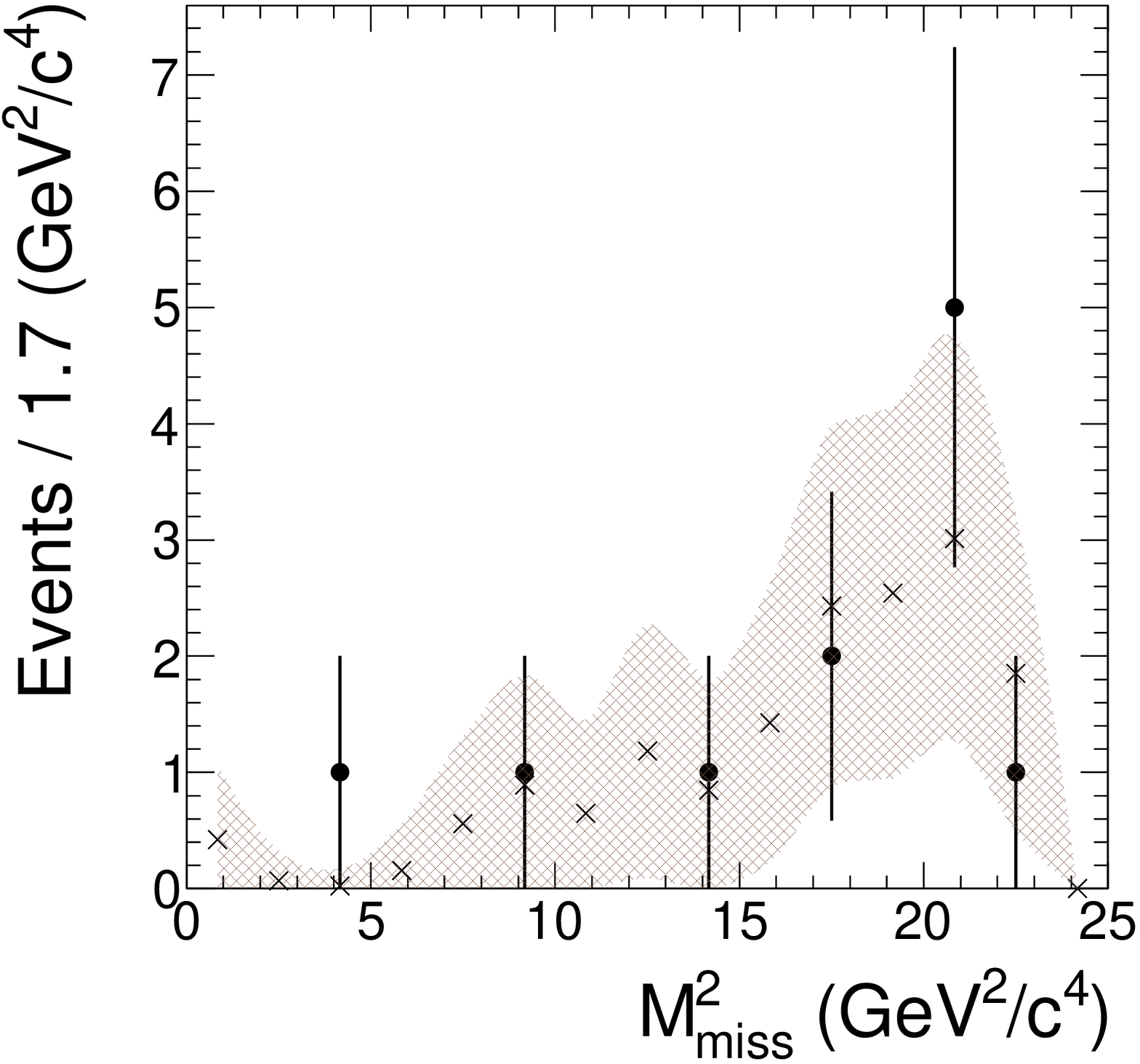} 
	\includegraphics[width=0.235\textwidth]{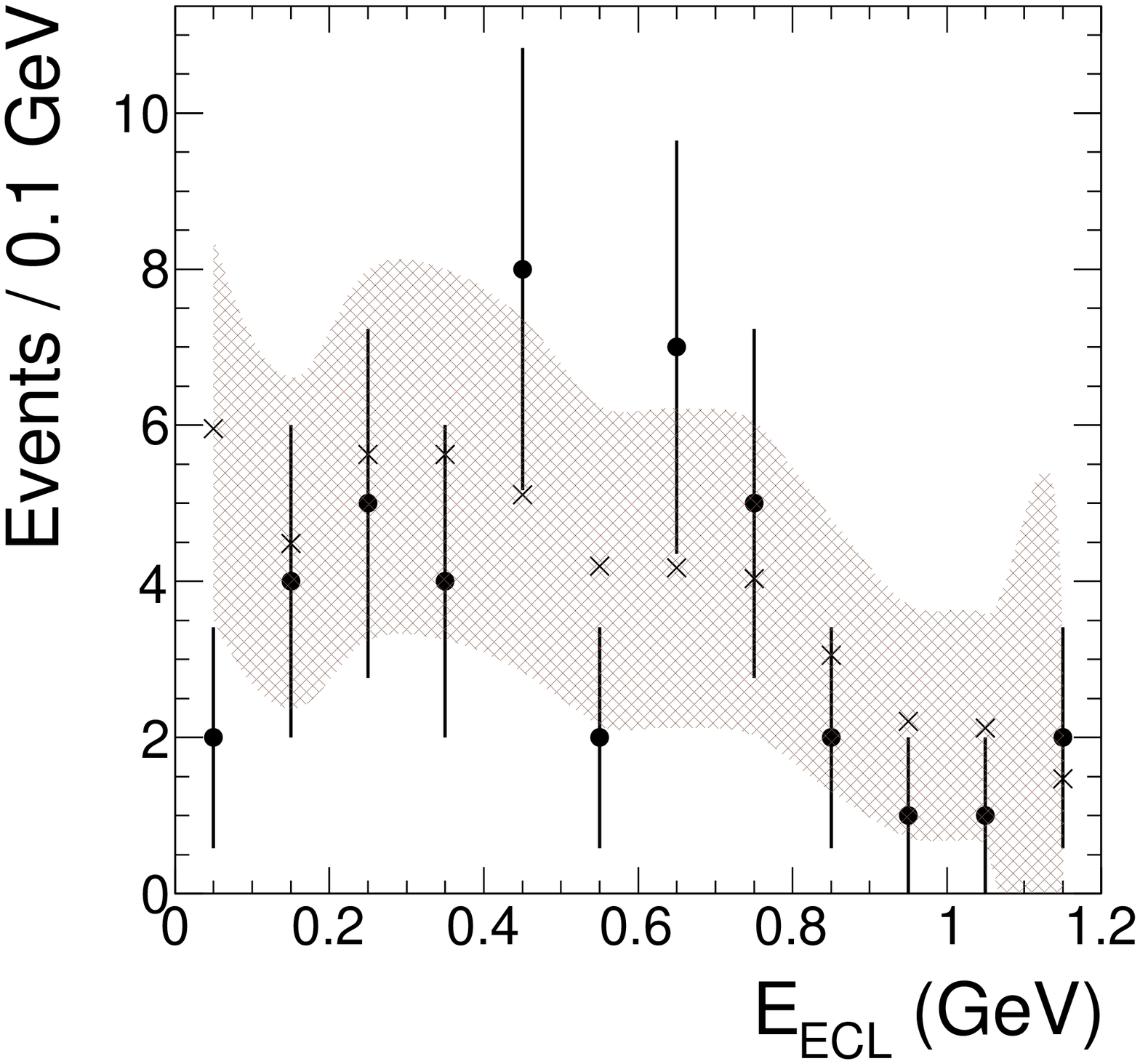} \caption{${M^2_\text{miss}}$ (left) and $E_\text{ECL}$ (right) distributions of data and the expected background for $B^0\to\text{invisible}+\gamma$. The ${M^2_\text{miss}}$ distribution is plotted in the $E_\text{ECL}$ signal box and the $E_\text{ECL}$ distribution is plotted in the whole ${M^2_\text{miss}}$ region. The black points with error are data. The gray crosses with a shaded error band are the expected background.}
	\label{fig7}
    \end{figure}

  Uncertainties of $N_\text{bkg, box}^{\text{\,data}}$ come from the statistical error of the first term and the systematic error of the second term in the right-hand side of Eq.~\ref{ratio}. The latter is estimated by a control sample $B^0\to D^-l^+\nu\,(l=e,\mu,D^-\to K^+ \pi^-\pi^-$). Similar to the case of $B^0\to D^{*-}l^+\nu$ , the double tagging, $M_{D^-}$ requirements, extra particles vetoes, $O_\text{tag}$ and $O_\text{shape}$ thresholds are applied. In the control sample, background numbers in the $E_{\text{ECL}}$ signal box and sideband are obtained through fitting the $M^2_\text{miss}$ distribution to data, which is shown in Fig.~\ref{fig6}. The ratio of the background yields in the two regions is compared with the ratio in the control sample MC simulation. The difference and the statistical uncertainty of fitting, which is between 16---20\%, are added in quadrature and taken as the systematic uncertainty. For $B^0\to\text{invisible}+\gamma$, the uncertainty is 33\% and for the binned cases, the uncertainties are between 23---30\%. The counting results in the $E_\text{ECL}$ signal box are shown for $B^0\to\text{invisible}+\gamma$ and the binned analysis in Table~\ref{Table3}. Figure~\ref{fig7} shows the ${M^2_\text{miss}}$ and $E_\text{ECL}$ distributions of data and the expected background for $B^0\to\text{invisible}+\gamma$. The observed numbers of events are all consistent within uncertainties with the expected backgrounds. \par
  
  	\begin{table}[htb]
 	\caption{Estimated number of background events in the signal box and the number of events in the signal box ($N_\text{box}^{\text{data}}$) for $B^0\to\text{invisible}+\gamma$ and ${M^2_\text{miss}}$ bins.} 		
	\begin{tabular}
	{@{\hspace{0.4cm}}l@{\hspace{0.4cm}}  @{\hspace{0.4cm}}c@{\hspace{0.4cm}}c@{\hspace{0.4cm}}}
	\hline\hline
	& $N_\text{bkg,box}^{\text{\,data}}$ & $N_\text{box}^{\text{data}}$\\ 
	\hline 
	$B^0\to\text{invisible}+\gamma$&$16.1\pm6.3$&$11$\\
	\hline
	bin1&$3.2\pm2.1$&$2$\\
	bin2&$1.0\pm0.8$&$2$\\
	bin3& $4.4\pm2.6$&$3$\\
	bin4 &$7.1\pm2.9$&$4$\\
	bin5 & $6.6\pm2.9$&$7$\\
	\hline\hline
   \end{tabular}
   \label{Table3}	
   \end{table}

Taking the data-MC difference in selection rates into account, the signal efficiencies are calibrated through the formula:
 \begin{equation}
 \epsilon_\text{sig}^{\text{data}} = \epsilon_\text{sig}^{\text{MC}} \times C_\text{FR}\times C_\text{tr}\times C_{\pi^0}\times C_{K_L^0}\times C_\text{NN} \\
  \end{equation} 
  where $\epsilon_\text{sig}^{\text{data}}$ and $\epsilon_\text{sig}^{\text{MC}}$ are the signal efficiencies from data and MC, respectively, $C_\text{FR}$, $C_\text{tr}$, $ C_{\pi^0}$, $C_{K_L^0}$ and $C_\text{NN}$ are calibration factors due to the full reconstruction process, the extra tracks, $\pi^0$, $K_L^0$ vetoes and the NN output thresholds, respectively. 
  The $C_\text{FR}$ factor has been studied~\cite{18} using charmed semileptonic signal-side $B$ decays, and its value depends on the $P_\text{FR}$ of the $B_\text{tag}$ and the tag-side reconstructed channel. For $B^0\to\text{invisible}$,  $B^0\to\text{invisible}+\gamma$, and the binned analysis, the $C_\text{FR}$ factor lies between 0.64 to 0.70. On the other hand, $C_\text{tr}$, $ C_{\pi^0}$, $C_{K_L^0}$, and $C_\text{NN}$ are estimated through control samples, in which the signal efficiencies before and after each selection on data and MC simulation are compared. The control samples include six modes, with the signal side decaying respectively through: $B^0\to D^{*-}l^+\nu$ ($D^{*-}\to \bar{D^0}\pi^-$, $\bar{D^0}\to K^+ \pi^-$); $B^-\to D^{*0}l^-\nu$ ($D^{*0}\to D^0\pi^0$, $D^0\to K^- \pi^+$); $B^0\to D^-l^+\nu$ ($D^-\to K^+ \pi^-\pi^-$), where $l=e$ or $\mu$. The events are doubly tagged with the selections on $M_{D^0}$, $M^2_\text{miss}$, and $\Delta M_D$ the same as mentioned before. In addition, we require $-0.5\text{ GeV}^2/c^4<M^2_\text{miss}<0$ ($0.4 \sigma$ window) for $B^0\to D^{-}l^+\nu $, $E_\text{ECL} <0.4 \text{ GeV}$ for all the control sample modes and the difference between the reconstructed $D^{*0}$ and $D^0$ masses to lie within $0.138 \text{ GeV}/c^2$ to $0.146 \text{ GeV}/c^2$ ($2.4 \sigma$ window) for $B^-\to D^{*0}l^-\nu$. The averaged calibration factors obtained from the six modes are used to calibrate the $B^0\to\text{invisible}\,(+\gamma)$ signal efficiencies. Results for $C_\text{tr}$, $ C_{\pi^0}$, and $C_{K_L^0}$ are 0.98, 0.96, and 1.06, respectively. For the $C_\text{NN}$, values vary between 0.90 to 0.95 according to the different $O_\text{tag}$ and $O_\text{shape}$ thresholds for $B^0\to\text{invisible}$,  $B^0\to\text{invisible}+\gamma$, and the binned analysis.\par 
  
  Systematic uncertainties associated with the signal efficiency are from the full reconstruction and signal-side selections. Uncertainties of the calibration factors contribute to both sources, which are 4.5\%, 3.0\%, 3.6\%, 3.2\% and 3.1\% for the full reconstruction, extra tracks, $\pi^0$, $K_L^0$ veto, and the NN output thresholds, respectively. For the modes with a photon, the uncertainties due to photon detection efficiency are within 2.8---3.0\%, which is studied using a radiative Bhabha sample and $B^0\to K^{*0}\gamma$ in the ECL barrel and endcap region, respectively~\cite{19}. Combining all the sources, the systematic uncertainty of the signal efficiency is 7.9\% for $B^0\to\text{invisible}$, and around 8.4\% for $B^0\to\text{invisible}+\gamma$ and the binned analysis. The calibrated signal efficiencies for $B^0\to\text{invisible}$ in the whole fitting region,  $B^0\to\text{invisible}+\gamma$ and the five bins in the $E_\text{ECL}$ signal box are $(7.1\pm0.6)\times10^{-4}$, $(5.5\pm0.5)\times10^{-4}$, $(6.3\pm0.5)\times10^{-4}$, $(7.7\pm0.6)\times10^{-4}$, $(6.6\pm0.5)\times10^{-4}$,
 $(7.2\pm0.6)\times10^{-4}$ and $(3.4\pm0.3)\times10^{-4}$, respectively.
 \par
  
 Since the signal yield is not significant for both  $B^0\to\text{invisible}$ and $B^0\to\text{invisible}+\,\gamma$ (whole range or the five $M_\text{miss}^2$ bins), upper limits at $90\%$ confidence level on the branching fraction ($\mathcal{B}_{UL}$) are calculated. For $B^0\to\text{invisible}$, the  upper limit is obtained by solving the equation:
 \begin{equation}
 \int_{0}^{\mathcal{B}_{UL}} \mathcal{L}\,(\mathcal{B}) d\mathcal{B}=0.9\int_{0}^{\infty} \mathcal{L}\,(\mathcal{B})d\mathcal{B},
 \end{equation}
 where $\mathcal{B}$ is the assumed branching fraction, and $\mathcal{L}\,(\mathcal{B})$ is the corresponding maximized likelihood from the fit on data. The 1.4\% uncertainty on the number of produced $B$-meson pairs, systematic uncertainties of signal yield and efficiency are taken into consideration by convolving the likelihood function with a Gaussian function whose width equals the total systematic uncertainty. The result is:
 \begin{center}
 $\mathcal{B}\,$($B^0\to\text{invisible})< 7.8\times10^{-5}$ at 90\% C.L.
 \end{center}\par

 For $B^0\to\text{invisible}+\,\gamma$, a frequentist style limit evaluated in the \textsc{TRolke} package~\cite{20} is used to obtain upper limits on the branching fraction. The method is based on the profile likelihood with the uncertainties on background and signal efficiency taken into account. The upper limits of the branching fraction are shown in Table~\ref{Table4}.  \par
 
    \begin{table}[h!tb]
	\centering
	\caption{Branching-fraction upper limits for the $B^0\to\text{invisible}+\gamma$ mode.}	
	\begin{tabular} {@{\hspace{0.5cm}}l@{\hspace{0.5cm}}  @{\hspace{0.5cm}}c@{\hspace{0.5cm}}}
	\hline \hline
	Channel& $\mathcal{B}$\\
	\hline
	$B^0\to\text{invisible}+\gamma$&$ < 1.6\times10^{-5}$\\
	\hline
	$B^0\to\text{invisible}+\gamma$, bin1&$ < 7.0\times10^{-6}$\\
	$B^0\to\text{invisible}+\gamma$, bin2&$ < 7.6\times10^{-6}$\\
	
	$B^0\to\text{invisible}+\gamma$, bin3&$ < 8.1\times10^{-6}$\\
	$B^0\to\text{invisible}+\gamma$, bin4&$< 5.4\times10^{-6}$\\
	$B^0\to\text{invisible}+\gamma$, bin5&$ < 2.8\times10^{-5}$\\
	\hline\hline 
    \end{tabular}		
	\label{Table4}			
    \end{table}

In summary, we have searched for the decays $B^0\to\text{invisible}$ and $B^0\to\text{invisible}+\gamma$ and find no evidence for them.  For the latter decay, the energy of the photon is required to be greater than 0.5 GeV. We set upper limits on the branching fractions $\mathcal{B}\,(B^0\to\text{invisible}) < 7.8\times10^{-5}$ and  $\mathcal{B}\,(B^0\to\text{invisible}+\gamma) < 1.6\times10^{-5}$ at $90\%$ confidence level. We improve upon the previous Belle limit~\cite{07} on $B^0\to\text{invisible}$, and the limit obtained for $B^0\to\text{invisible}+\gamma$ is the most stringent.   \par

We thank the KEKB group for the excellent operation of the
accelerator; the KEK cryogenics group for the efficient
operation of the solenoid; and the KEK computer group, and the Pacific Northwest National
Laboratory (PNNL) Environmental Molecular Sciences Laboratory (EMSL)
computing group for strong computing support; and the National
Institute of Informatics, and Science Information NETwork 5 (SINET5) for
valuable network support.  We acknowledge support from
the Ministry of Education, Culture, Sports, Science, and
Technology (MEXT) of Japan, the Japan Society for the 
Promotion of Science (JSPS), and the Tau-Lepton Physics 
Research Center of Nagoya University; 
the Australian Research Council including grants
DP180102629, 
DP170102389, 
DP170102204, 
DP150103061, 
FT130100303; 
Austrian Science Fund (FWF);
the National Natural Science Foundation of China under Contracts
No.~11435013,  
No.~11475187,  
No.~11521505,  
No.~11575017,  
No.~11675166,  
No.~11705209;  
Key Research Program of Frontier Sciences, Chinese Academy of Sciences (CAS), Grant No.~QYZDJ-SSW-SLH011; 
the  CAS Center for Excellence in Particle Physics (CCEPP); 
the Shanghai Pujiang Program under Grant No.~18PJ1401000;  
the Ministry of Education, Youth and Sports of the Czech
Republic under Contract No.~LTT17020;
the Carl Zeiss Foundation, the Deutsche Forschungsgemeinschaft, the
Excellence Cluster Universe, and the VolkswagenStiftung;
the Department of Science and Technology of India; 
the Istituto Nazionale di Fisica Nucleare of Italy; 
National Research Foundation (NRF) of Korea Grant
Nos.~2016R1\-D1A1B\-01010135, 2016R1\-D1A1B\-02012900, 2018R1\-A2B\-3003643, 2018R1\-A4A\-1025334, 2018R1\-A6A1A\-06024970, 2018R1\-D1A1B\-07047294, 2019K1\-A3A7A\-09033840,
2019R1\-I1A3A\-01058933;
Radiation Science Research Institute, Foreign Large-size Research Facility Application Supporting project, the Global Science Experimental Data Hub Center of the Korea Institute of Science and Technology Information and KREONET/GLORIAD;
the Polish Ministry of Science and Higher Education and 
the National Science Center;
the Ministry of Science and Higher Education of the Russian Federation, Agreement 14.W03.31.0026; 
University of Tabuk research grants
S-1440-0321, S-0256-1438, and S-0280-1439 (Saudi Arabia);
the Slovenian Research Agency;
Ikerbasque, Basque Foundation for Science, Spain;
the Swiss National Science Foundation; 
the Ministry of Education and the Ministry of Science and Technology of Taiwan;
and the United States Department of Energy and the National Science Foundation.

\end{document}

%% file: pub554_authorlist.tex
\noaffiliation
\affiliation{University of the Basque Country UPV/EHU, 48080 Bilbao}
\affiliation{Beihang University, Beijing 100191}
\affiliation{Brookhaven National Laboratory, Upton, New York 11973}
\affiliation{Budker Institute of Nuclear Physics SB RAS, Novosibirsk 630090}
\affiliation{Faculty of Mathematics and Physics, Charles University, 121 16 Prague}
\affiliation{University of Cincinnati, Cincinnati, Ohio 45221}
\affiliation{Deutsches Elektronen--Synchrotron, 22607 Hamburg}
\affiliation{Department of Physics, Fu Jen Catholic University, Taipei 24205}
\affiliation{Key Laboratory of Nuclear Physics and Ion-beam Application (MOE) and Institute of Modern Physics, Fudan University, Shanghai 200443}
\affiliation{Gifu University, Gifu 501-1193}
\affiliation{II. Physikalisches Institut, Georg-August-Universit\"at G\"ottingen, 37073 G\"ottingen}
\affiliation{SOKENDAI (The Graduate University for Advanced Studies), Hayama 240-0193}
\affiliation{Gyeongsang National University, Jinju 52828}
\affiliation{Department of Physics and Institute of Natural Sciences, Hanyang University, Seoul 04763}
\affiliation{University of Hawaii, Honolulu, Hawaii 96822}
\affiliation{High Energy Accelerator Research Organization (KEK), Tsukuba 305-0801}
\affiliation{J-PARC Branch, KEK Theory Center, High Energy Accelerator Research Organization (KEK), Tsukuba 305-0801}
\affiliation{Higher School of Economics (HSE), Moscow 101000}
\affiliation{IKERBASQUE, Basque Foundation for Science, 48013 Bilbao}
\affiliation{Indian Institute of Science Education and Research Mohali, SAS Nagar, 140306}
\affiliation{Indian Institute of Technology Bhubaneswar, Satya Nagar 751007}
\affiliation{Indian Institute of Technology Madras, Chennai 600036}
\affiliation{Institute of High Energy Physics, Chinese Academy of Sciences, Beijing 100049}
\affiliation{Institute of High Energy Physics, Vienna 1050}
\affiliation{Institute for High Energy Physics, Protvino 142281}
\affiliation{INFN - Sezione di Napoli, 80126 Napoli}
\affiliation{INFN - Sezione di Torino, 10125 Torino}
\affiliation{Advanced Science Research Center, Japan Atomic Energy Agency, Naka 319-1195}
\affiliation{J. Stefan Institute, 1000 Ljubljana}
\affiliation{Institut f\"ur Experimentelle Teilchenphysik, Karlsruher Institut f\"ur Technologie, 76131 Karlsruhe}
\affiliation{Kavli Institute for the Physics and Mathematics of the Universe (WPI), University of Tokyo, Kashiwa 277-8583}
\affiliation{Kennesaw State University, Kennesaw, Georgia 30144}
\affiliation{Kitasato University, Sagamihara 252-0373}
\affiliation{Korea Institute of Science and Technology Information, Daejeon 34141}
\affiliation{Korea University, Seoul 02841}
\affiliation{Kyungpook National University, Daegu 41566}
\affiliation{LAL, Univ. Paris-Sud, CNRS/IN2P3, Universit\'{e} Paris-Saclay, Orsay 91898}
\affiliation{P.N. Lebedev Physical Institute of the Russian Academy of Sciences, Moscow 119991}
\affiliation{Faculty of Mathematics and Physics, University of Ljubljana, 1000 Ljubljana}
\affiliation{Ludwig Maximilians University, 80539 Munich}
\affiliation{Luther College, Decorah, Iowa 52101}
\affiliation{University of Maribor, 2000 Maribor}
\affiliation{Max-Planck-Institut f\"ur Physik, 80805 M\"unchen}
\affiliation{School of Physics, University of Melbourne, Victoria 3010}
\affiliation{University of Mississippi, University, Mississippi 38677}
\affiliation{Moscow Physical Engineering Institute, Moscow 115409}
\affiliation{Graduate School of Science, Nagoya University, Nagoya 464-8602}
\affiliation{Universit\`{a} di Napoli Federico II, 80055 Napoli}
\affiliation{Nara Women's University, Nara 630-8506}
\affiliation{National United University, Miao Li 36003}
\affiliation{Department of Physics, National Taiwan University, Taipei 10617}
\affiliation{H. Niewodniczanski Institute of Nuclear Physics, Krakow 31-342}
\affiliation{Nippon Dental University, Niigata 951-8580}
\affiliation{Niigata University, Niigata 950-2181}
\affiliation{Novosibirsk State University, Novosibirsk 630090}
\affiliation{Osaka City University, Osaka 558-8585}
\affiliation{Pacific Northwest National Laboratory, Richland, Washington 99352}
\affiliation{Panjab University, Chandigarh 160014}
\affiliation{Peking University, Beijing 100871}
\affiliation{University of Pittsburgh, Pittsburgh, Pennsylvania 15260}
\affiliation{Punjab Agricultural University, Ludhiana 141004}
\affiliation{Theoretical Research Division, Nishina Center, RIKEN, Saitama 351-0198}
\affiliation{University of Science and Technology of China, Hefei 230026}
\affiliation{Seoul National University, Seoul 08826}
\affiliation{Showa Pharmaceutical University, Tokyo 194-8543}
\affiliation{Soochow University, Suzhou 215006}
\affiliation{Soongsil University, Seoul 06978}
\affiliation{Sungkyunkwan University, Suwon 16419}
\affiliation{School of Physics, University of Sydney, New South Wales 2006}
\affiliation{Department of Physics, Faculty of Science, University of Tabuk, Tabuk 71451}
\affiliation{Tata Institute of Fundamental Research, Mumbai 400005}
\affiliation{School of Physics and Astronomy, Tel Aviv University, Tel Aviv 69978}
\affiliation{Toho University, Funabashi 274-8510}
\affiliation{Department of Physics, Tohoku University, Sendai 980-8578}
\affiliation{Department of Physics, University of Tokyo, Tokyo 113-0033}
\affiliation{Tokyo Institute of Technology, Tokyo 152-8550}
\affiliation{Tokyo Metropolitan University, Tokyo 192-0397}
\affiliation{Virginia Polytechnic Institute and State University, Blacksburg, Virginia 24061}
\affiliation{Wayne State University, Detroit, Michigan 48202}
\affiliation{Yamagata University, Yamagata 990-8560}
\affiliation{Yonsei University, Seoul 03722}
  \author{Y.~Ku}\affiliation{Department of Physics, National Taiwan University, Taipei 10617} 
  \author{P.~Chang}\affiliation{Department of Physics, National Taiwan University, Taipei 10617} 
    
  \author{I.~Adachi}\affiliation{High Energy Accelerator Research Organization (KEK), Tsukuba 305-0801}\affiliation{SOKENDAI (The Graduate University for Advanced Studies), Hayama 240-0193} 
  \author{K.~Adamczyk}\affiliation{H. Niewodniczanski Institute of Nuclear Physics, Krakow 31-342} 
  \author{H.~Aihara}\affiliation{Department of Physics, University of Tokyo, Tokyo 113-0033} 
  \author{D.~M.~Asner}\affiliation{Brookhaven National Laboratory, Upton, New York 11973} 
  \author{T.~Aushev}\affiliation{Higher School of Economics (HSE), Moscow 101000} 
  \author{R.~Ayad}\affiliation{Department of Physics, Faculty of Science, University of Tabuk, Tabuk 71451} 
  \author{P.~Behera}\affiliation{Indian Institute of Technology Madras, Chennai 600036} 
  \author{C.~Bele\~{n}o}\affiliation{II. Physikalisches Institut, Georg-August-Universit\"at G\"ottingen, 37073 G\"ottingen} 
% ...
\collaboration{The Belle Collaboration}